\documentclass[a4paper,11pt]{article}
\pdfoutput=1 % if your are submitting a pdflatex (i.e. if you have
\usepackage{jheppub} % for details on the use of the package, please
\usepackage[T1]{fontenc} % if needed

\title{Crystallographic Interacting Topological Phases and Equivariant Cohomology:\\ To assume or not to assume}% Force line breaks with \\
\author[a,1]{Daniel Sheinbaum,\note{Corresponding author.}}
\author[b,2]{Omar Antol\'in Camarena}

\affiliation[a]{Facultad de Qu\'imica, Universidad Nacional Aut\'onoma de M\'exico, 04510  Mexico City, Mexico.}
\affiliation[b]{Instituto de Matematicas, Universidad Nacional Aut\'onoma de M\'exico, 04510  Mexico City, Mexico.}%\\This line break forced with \textbackslash\textbackslash
\emailAdd{dshein@ciencias.unam.mx}
\emailAdd{omar@matem.unam.mx}

\abstract{For symmorphic crystalline interacting gapped systems we derive a classification under adiabatic evolution. This classification is complete for non-degenerate ground states. For the degenerate case we discuss some invariants given by equivariant characteristic classes. We do not assume an emergent relativistic field theory nor that phases form a topological spectrum. We also do not restrict to systems with short-range entanglement, stability against stacking with trivial systems nor assume the existence of quasi-particles as is done in SPT and SET classifications respectively. Using a slightly generalized Bloch decomposition and Grassmanians made out of ground state spaces, we show that the $P$-equivariant cohomology of a $d$-dimensional torus gives rise to different interacting phases, where $P$ denotes the point group of the crystalline structure. We compare our results to bosonic symmorphic crystallographic SPT phases and to non-interacting fermionic crystallographic phases in class A. Finally we discuss the relation of our assumptions to those made for crystallographic SPT and SET phases.}

\newcommand{\Hol}{\mathsf{Hol}}

\begin{document} 
%\maketitle
\flushbottom


%Use showkeys class option if keyword
\maketitle

\section{Introduction}\label{sec:Introduction}

The role of algebraic topology in the classification of topological phases of matter began with the pioneering work of Thouless et al. \cite{TKNN} and Avron et al. \cite{Avron-Seiler-Simon} on the integer quantum Hall effect in the 80's. There was a new surge of interest in topological phases due to the discovery of topological insulators \cite{Kane-Mele} in the mid 2000's, culminating in the classification for non-interacting fermion systems using characteristic classes \cite{D-G-AI}, \cite{D-G-AII} and equivariant $K$-theory \cite{Freed-Moore},\cite{Kitaev} which still appears even in the presence of disorder \cite{Prodan-Schulz-Baldes}, \cite{Thiang1}.

Interacting topological phases, on the other hand, have been much harder to analyse, beginning with the work on the fractional quantum Hall effect \cite{Laughlin-FQHE}, \cite{Wilczek-statistics} and the concept of topological order \cite{Wen-Top-Order}. For symmetry protected (SPT) topological phases there are many constructive classifications \cite{Wen-SPT}, \cite{Kapustin-SPT}, \cite{Freed-SPT-I}, \cite{Debray-Fermionic}, \cite{Debray-Semion}, \cite{Gaiotto-SPT}, \cite{Thorngren-SET} to mention the trendy ones. We refer to these as constructive because they do not derive their classification from first principles, but rather construct their classification either by assuming that there is an effective relativistic extended topological field theory emerging at low energy, or that the set of short-range entangled invertible phases in all dimensions conspire to form a spectrum in the sense of algebraic topology, and there are different possible choices of spectra competing out there in the literature (see \cite{Xiong-SPT} for a discussion).  We will instead derive a classification for symmorphic crystallographic interacting systems under adiabatic evolution \cite{Avron-Adiabatic}, \cite{Sven-Adiabatic} without these added assumptions: neither restricting to short-range entangled systems, nor declaring that phases are stable under stacking with a system in a trivial phase. Using a modest generalization of the Bloch decomposition to the many-body case, we show that different equivariant characteristic classes \cite{Bott-Tu} in the cohomology ring $H^{*}_{P}(\mathbb{T}^d;\mathbb{Z})$ give rise to different interacting topological phases for degenerate and non-degenerate ground states. For the non-degenerate case we provide a full classification for symmorphic crystallographic groups given by $H^{2}_P(\mathbb{T}^d;\mathbb{Z})$ and compare our results in $d=2$ with those of Thorngren and Else \cite{Thorngren-SET} for the analogue SPT systems. Here $P$ denotes the point group of the crystalline structure. We find that every representative of these non-degenerate phases can be adiabatically connected to a non-interacting system with a single valence band. We also compare, in the simplest cases, our non-degenerate fermionic classification to the non-interacting fermionic gapped phases of Shiozaki et al. \cite{Gomi-Wallpaper} classified by equivariant $K$-theory. Let us mention from the outset that our results do not imply that the published SPT classifications are incorrect as we do not have all of the same starting assumptions. Our calculations can be seen as measuring the effect of removing the assumptions mentioned previously. We discuss in section \ref{sec:Difference} some plausible scenarios. Symmetry enriched topological phases (SET) \cite{Mesaros-SET}, \cite{Cheng-SET}, \cite{Vishwanath-SET} are traditionally classified by their fractionalized quasi-particle excitations (sometimes arising from an emergent relativistic field theory). The classification method for these is completely different from the ones for SPTs. Our derivation applies to both the degenerate and non-degenerate ground state of symmorphic crystalline systems in a unifying manner (but it should be noted that in the degenerate case, the cohomological invariants which are well understood in the algebraic topology literature only allow a partial classification of these phases). However we also do not have the same assumptions as the SET classifications since we do not assume (and hence do not use) the existence of quasi-particle excitations nor an algebraic structure associated to these. 

\section{Translation and Fock space decomposition}\label{sec:Translation}

Common physical lore states that there is no Brillouin zone and no generalized single particle picture for interacting systems. We now explain that the same mathematical construction of the Brillouin zone used for non-interacting systems can also be used for interacting systems. Consider for continuous systems the one-particle Hilbert space $L^2(\mathbb{R}^{d};W)$, with $W$ an $n$-dimensional representation of $\mathfrak{su}(2)$ corresponding to the spin of the particle. We can construct a Fock space $\mathfrak{F}(L^2(\mathbb{R}^{d};W)) = \bigoplus_{n\geq 0}\,L^2(\mathbb{R}^{d};W)^{\otimes n}$ and there are bosonic and fermionic Fock spaces $\mathfrak{F}_{+}(L^2(\mathbb{R}^{d};W)) = \bigoplus_{n\geq 0} \mathrm{Sym}^{n}(L^2(\mathbb{R}^{d};W))$ and $\mathfrak{F}_{-}(L^2(\mathbb{R}^{d};W)) = \bigoplus_{n\geq 0}\Lambda^{n}(L^2(\mathbb{R}^{d};W))$ made out of symmetric and antisymmetric wave functions. Let $\mathfrak{H}^{n}$ denote either $\mathrm{Sym}^{n}(L^2(\mathbb{R}^{d};W))$ or $\Lambda^{n}(L^2(\mathbb{R}^{d};W))$. Since $\mathbb{Z}^{d}$ acts on $\mathbb{R}^{d}$ by translations, it also acts on $L^2(\mathbb{R}^{d};W)$, and therefore on $\mathfrak{H}^{n}$ diagonally by acting on each factor. Just as in the single particle case ($n = 1$) we can perform a Bloch decomposition \cite{Gomi-Twists} using the dual group $\widehat{\mathbb{Z}}^{d} = \mathrm{Hom}(\mathbb{Z}^{d},U(1))$ and write it as a \textit{direct integral decomposition}:
\begin{equation}
 \mathfrak{H}^{n} \simeq \int^{\oplus}_{\widehat{\mathbb{Z}}^{d}} \mathfrak{H}^{n}(\vec{k})d\vec{k}\,.
\end{equation}

This decomposition can be intuitively thought of as a direct sum of Hilbert spaces, one for each point of $\widehat{\mathbb{Z}}^{d} \cong \mathbb{T}^{d}$, a $d$-dimensional torus. We call this torus a generalized Brillouin zone, since, though mathematically it is obtained in the same way, its physical interpretation is rather different as $\vec{k}$ is a many-body crystal momentum. The spaces $\mathfrak{H}^{n}(\vec{k})$ are easily defined:
\begin{equation}
    \mathfrak{H}^{n}(\vec{k}) = \{ \psi \in \mathfrak{H}^{n}|\, T_{\vec{a}}\psi = e^{i\vec{k}\cdot\vec{a}}\psi \,\,\forall \vec{a}\in \mathbb{Z}^{d}\}
\end{equation}
Obtaining this decomposition is as simple as for the single particle case, the only differences being in a few details presented in Appendix \ref{sec:TranslationA}. Note that this is simply a rewriting of the Hilbert space and is completely independent of the Hamiltonian. Now we can repeat the same process for any $n$ and hence decompose the entire Fock space as
\begin{equation}\label{eq:Fock-direct}
    \mathfrak{F}_{\pm}(L^2(\mathbb{R}^{d};W))\simeq \int^{\oplus}_{\mathbb{T}^{d}} \bigoplus_{n\geq 0}\mathfrak{H}^{n}(\vec{k})d\vec{k}\,.
\end{equation}

where, just as one can swap integrals with sums, one can do the same for direct integrals and direct sums. This rewriting of Fock space by itself is not useful; it becomes useful when a many-body Hamiltonian operator $\mathcal{H}$ has interaction terms which are discrete translation invariant (e.g. $\mathcal{H} = \sum_{j}(i\nabla_{j} -e\vec{A}(x_{j}))^2 + V(x_{j}) +\sum_{i\neq j}e^2/|x_{i}-x_{j}|$), with $\vec{A}(x_{j}),\,V(x_{j})$ the periodic and magnetic potentials associated to the ions and hence it commutes with the unitary representation $U$ of $\mathbb{Z}^{d}$ on $\mathfrak{F}_{\pm}(L^2(\mathbb{R}^{d};W))$, that is $[\mathcal{H},U(\gamma)] = 0$ for all $\gamma \in \mathbb{Z}^d$. In that case $\mathcal{H}$ can be decomposed (``diagonalized'') into a family of operators parametrized by $\mathbb{T}^{d}$ and written as:
\begin{equation}
    \mathcal{H}\simeq \int^{\oplus}_{\mathbb{T}^{d}} \mathcal{H}(\vec{k})d\vec{k}.
\end{equation}
The spectrum $\sigma(\mathcal{H}(\vec{k}))$ varies continuously with $\vec{k}$.
From the analytical point of view we have not made much progress at solving the many-body problem as the spectral subspaces of $\mathcal{H}(\vec{k})$ are still many-body wave functions and $\vec{k}$ is a quantum number behaving like a total momentum of all particles combined. Perhaps this is why this decomposition had been completely dismissed. However, from the topological point of view it is a big gain as we now have the topological space $\mathbb{T}^{d}$ to work with.

\section{Gapped phases and Grassmanians}\label{sec:gapped}

We now assume that the spectrum of the Hamiltonian is gapped: that for some $\delta > 0$, $\sigma(\mathcal{H})\cap (0,\delta) = \emptyset$. Note that there are many examples of Hamiltonians which are not gapped so this is a strong restriction already. Also note that the gapped condition is physically meaningful only in the thermodynamic limit. General theorems \cite{Reed-Simon-IV} for direct integral decompositions imply that if a translation invariant $\mathcal{H}$ is gapped then $\mathcal{H}(\vec{k})$ is also gapped except at most for a measure zero set in $\mathbb{T}^{d}$, and such cases are non-generic so without loss of generality we can further restrict to Hamiltonians $\mathcal{H}$ such that $\sigma(\mathcal{H}(\vec{k}))\cap (0,\delta) = \emptyset$ for all $\vec{k}$. Thus for each $\vec{k}$, $\mathcal{H}(\vec{k})$ singles out a subspace $\mathfrak{G}(\vec{k})$ of $\mathfrak{F}_\pm(\vec{k}) := \bigoplus_{n\geq 0}\mathfrak{H}^n(\vec{k})$ corresponding to the ground state space. Let $m$ be the ground state degeneracy, so $\dim \mathfrak{G}(\vec{k}) = m$. The set of all $m$-dimensional subspaces of $\mathfrak{F}_\pm(\vec{k})$ can be given a topology making it into a space called a Grassmanian \cite{Grassmannians}, denoted by $\mathcal{G}_m(\mathfrak{F}_\pm(\vec{k}))$ and $\mathfrak{G}(\vec{k})$ is a point in this space.

By Kuiper's theorem \cite{Kuiper}, any bundle of Hilbert spaces is trivial, so the bundle $\mathfrak{F}_\pm(\vec{k})$ is isomorphic to one with no dependence on $\vec{k}$, and thus the $\mathfrak{F}_\pm(\vec{k})$ can be replaced by a fixed Hilbert space $\mathfrak{H}^{'}$. We then obtain a continuous function $\mathbb{T}^{d} \rightarrow \mathcal{G}_{m}(\mathfrak{H}^{'})$ given by $\vec{k} \mapsto \mathfrak{G}(\vec{k})$. Here we have, in a sense, a proto-band theory: a single many-body band corresponding to $\mathfrak{G}(\vec{k})$; however we believe that here is where the resemblance to semi-classical conduction theory \cite{Ashcroft-Mermin} ends. 

Two systems are said to be in the same phase if one can adiabatically evolve one into the other \cite{Avron-Adiabatic}, \cite{Sven-Adiabatic}. Here, following \cite{Avron-Seiler-Simon}, we shall model adiabatic evolution as a continuous path $\mathcal{H}(s), \,s\in [0,1]$ of gapped Hamiltonians such that $\mathcal{H}(0) = \mathcal{H}_{0}$ and $\mathcal{H}(1) = \mathcal{H}_{1}$. This homotopy model has limitations which have been discussed in \cite{Sven-Adiabatic}, \cite{Sven-Ogata}. Also note that published SPT classifications only look at adiabatic paths which come from a local unitary evolution defined by Chen, Gu and Wen \cite{Wen-LUE}. We do not restrict ourselves to such paths (see Appendix \ref{sec:LUE} for a discussion). Thus, we are interested in homotopy classes of ground state space maps $\mathbb{T}^{d} \to \mathcal{G}_{m}(\mathfrak{H}')$:
\begin{equation}
    [\mathbb{T}^{d}, \mathcal{G}_{m}(\mathfrak{H}')] \cong \mathsf{Vect}^{m}(\mathbb{T}^{d}),
\end{equation}
where $\mathsf{Vect}^{m}(\mathbb{T}^{d})$ is the set of isomorphism classes of $m$-dimensional vector bundles on the torus. This last equivalence comes from the fact that the homotopy type of $\mathcal{G}_{m}(\mathfrak{H}')$ is that of the classifying space for vector bundles, $BGL(m)$ \cite{Grassmannians}. This means that we can associate to each Hamiltonian an $m$-dimensional vector bundle made out of the ground state space over the generalized Brillouin zone. Thus far we have only enforced that our Hamiltonians be gapped and $\mathbb{Z}^{d}$-invariant, is there any difference at this level between bosons and fermions? No. The homotopy type of $\mathcal{G}_{m}(\mathfrak{H}')$ by definition does not care if the Hilbert space inside the parenthesis is fermionic, bosonic, one particle or many particle, so long as its complex and has a countable orthonormal basis \cite{Grassmannians} which all of the above fulfil. So, is there any way to differentiate between bosons and fermions? The distinction is captured by further imposing on physical grounds that Hamiltonians of fermionic systems must have fermion parity symmetry $(-1)^{F}$ i.e. $[\mathcal{H},(-1)^{F}]= 0$, and we have
\begin{equation}
    [\mathbb{T}^{d}, \mathcal{G}_{m}(\mathfrak{H}')]_{\mathbb{Z}^{F}_2} \cong \mathsf{Vect}^{m}_{\mathbb{Z}^{F}_{2}}(\mathbb{T}^{d}),
\end{equation}
where $\mathbb{Z}^{F}_{2}$ is generated by $(-1)^{F}$, which acts trivially on $\mathbb{T}^{d}$ but not on $\mathcal{G}_{m}(\mathfrak{H}')$; and $\mathsf{Vect}^m_{\mathbb{Z}^F_2}(\mathbb{T}^d)$ denotes isomorphism classes of $\mathbb{Z}^F_2$-equivariant vector bundles on the torus. Fermion parity plays a more prominent role when combined with time-reversal and charge conjugation symmetry but we shall leave such combinations for future work.

\section{Crystallographic Interacting Phases}

Let us now assume that our Hamiltonian commutes with a crystallographic symmetry group $G$, containing $\mathbb{Z}^{d}$ and also the point group $P$, of allowed reflections and rotations. Here there are two important types of $G$, symmorphic and non-symmorphic \cite{Freed-Moore}, \cite{Gomi-Twists}. For now we shall only handle symmorphic groups, which means that $G$ is a semi-direct product $G = \mathbb{Z}^{d} \rtimes P$. In this case, there is a canonical $P$-action on the ground state bundle \cite{Gomi-Twists}. We are interested in adiabatic evolutions which preserve the extra symmetry given by $P$, so our phases will be described by
\begin{equation}\label{eq:Equivbosonic}
    [\mathbb{T}^{d}, \mathcal{G}_{m}(\mathfrak{H}')]_{P} \cong \mathsf{Vect}^{m}_{P}(\mathbb{T}^{d}).
\end{equation}
 For fermions we instead have to consider a generalized symmetry group $G_{F}$ defined via a group extension \cite{Adem-Groupcoho}, \cite{Thorngren-SET}
 \begin{equation}\label{eq:Fermionicsymmetry}
     1\rightarrow \mathbb{Z}_{2}^{F}\rightarrow G_{F} \rightarrow G\rightarrow 1,
 \end{equation}
 as these symmetries may combine non-trivially with fermion parity. Again, we only handle the symmorphic case in which $G_{F} = \mathbb{Z}^{d}\rtimes P_{F}$, where $P_F$ is given by an extension,
  \begin{equation}\label{eq:Fermionicpointsymmetry}
     1\rightarrow \mathbb{Z}_{2}^{F}\rightarrow P_{F} \rightarrow P\rightarrow 1.
 \end{equation}
Under these assumptions our phases will be given by:
\begin{equation}\label{eq:Equivfermionic}
    [\mathbb{T}^{d}, \mathcal{G}_{m}(\mathfrak{H}')]_{P_F} \cong \mathsf{Vect}^{m}_{P_F}(\mathbb{T}^{d}).
\end{equation}

\section{$m = 1$, the non-degenerate case}
 
 For $m=1$, the non-degenerate case, our Grassmanian is homotopy equivalent to $BU(1) \simeq \mathbb{C}P^{\infty}$, the infinite dimensional complex projective space which is an Eilenberg-Maclane space $K(\mathbb{Z},2)$, and is the classifying space of the ordinary cohomology group $H^2$ \cite{Hatcher-Alg-Top}. Hence for non-degenerate symmorphic crystallographic interacting phases we have (see Appendix \ref{sec:Borel} for details)
\begin{equation}
    [\mathbb{T}^d, \mathcal{G}_{1}(\mathfrak{H}')]_{P} \cong H^{2}_{P}(\mathbb{T}^d;\mathbb{Z})
\end{equation}
 for bosons and for fermions it is $H^{2}_{P_F}(\mathbb{T}^d;\mathbb{Z})$. We remark that this isomorphism is only as a set, as the group addition does not correspond to the so called \textit{stacking operation}. Nonetheless this outcome is sufficient to know all possible non-degenerate crystalline phases and to distinguish among them, and is, in that sense, complete. Ideally we would take as an example some many-body Hamiltonian with interaction terms and find the equivariant Chern classes of its ground state bundle. Unfortunately this would likely involve solving the interacting many-body problem and we were unable to find exact solutions in the literature to compute to which phase they belong to. We also note that the only method we are aware of for constructing interacting toy models, the so called Haldane's pseudo-potentials \cite{Seiringer-Pseudo-potentials} requires one to place the particles on a cylinder or sphere and it seems non-trivial to adapt our generalized Bloch decomposition to such cases. Instead here we start at the opposite end and present all possible bosonic phases for $m=1, d =2$ symmorphic wallpaper groups.\\
 To gain some perspective on the results presented below we note that this is analogous to employing the fundamental theorem of algebra to conclude how many roots a given polynomial has. The theorem does not tell one which numbers are roots of a given polynomial, it only tells us that there are $n$ for a degree $n$ polynomial. Similarly, our algebraic methods yield all symmorphic crystalline phases for $m =1$ stable under adiabatic evolution with interections that do not break symmetry, gap nor non-degeneracy but it does not say anything about which Hamiltonians are representatives of these phases. Nevertheless we show there are physical representatives of all of these phases in section \ref{sec:Examples}.\\ These groups were previously computed by Gomi \cite{Gomi-Twists}. To compare to the SPT literature, we picked the work of Thorngren and Else \cite{Thorngren-SET}, where they discuss how their approach considers both the group cohomology and cobordism crystallographic SPTs. They obtain in general the group $H^{2+d}(G;\mathbb{Z})$, corresponding to group cohomology \cite{Adem-Groupcoho} and $G = \mathbb{Z}^{d}\rtimes P$ is again the full crystallographic group.  In Table \ref{Table1} we compare to their $d=2$ case.
 
 \begin{table}[htp]
\begin{tabular}{cccc}
\hline
{\centering{$G$}} &  $P$ &       $H^2_P(\mathbb{T}^2;\mathbb{Z})$        & $H^4(G;\mathbb{Z})$    \\
\hline
        $\mathsf{p2}$   & $\mathbb{Z}_2$            & $\mathbb{Z}\oplus\mathbb{Z}_2^{3}$                     & $\mathbb{Z}_{2}^{4}$                   \\
  $\mathsf{p3}$  & $\mathbb{Z}_3$             & $\mathbb{Z}\oplus\mathbb{Z}_3^{2}$                     & $\mathbb{Z}_{3}^{3}$                   \\
$\mathsf{p4}$  & $\mathbb{Z}_4$             & $\mathbb{Z}\oplus\mathbb{Z}_2\oplus\mathbb{Z}_{4}$ & $\mathbb{Z}_2\oplus \mathbb{Z}_{4}^2$ \\
$\mathsf{p6}$   & $\mathbb{Z}_6$             & $\mathbb{Z}\oplus \mathbb{Z}_{6}$                     & $\mathbb{Z}_2^{2}\oplus\mathbb{Z}_{3}^2$                   \\
 $\mathsf{pm}$ & $\mathbb{Z}_2$             & $\mathbb{Z}_2^{2}$ & $\mathbb{Z}_{2}^2$ \\
 $\mathsf{cm}$  & $\mathbb{Z}_2$             & $\mathbb{Z}_2$ & $\mathbb{Z}_2$ \\
 $\mathsf{pmm}$  & $D_2$             & $\mathbb{Z}_2^4$ & $\mathbb{Z}_{2}^{8}$ \\
$\mathsf{cmm}$  & $D_2$            & $\mathbb{Z}_2^{3}$ & $\mathbb{Z}_{2}^5$ \\
 $\mathsf{p31m}$  & $D_3$            & $\mathbb{Z}_2$ & $\mathbb{Z}_{2}\oplus \mathbb{Z}_3$ \\
 $\mathsf{p3m1}$  & $D_3$            &  $\mathbb{Z}_2\oplus\mathbb{Z}_3$& $\mathbb{Z}_2$ \\
 $\mathsf{p4m}$  & $D_4$            & $\mathbb{Z}_2^{3}$ &
 $\mathbb{Z}_2^6$ \\
 $\mathsf{p6m}$  & $D_6$            & $\mathbb{Z}_2^{2}$ & $\mathbb{Z}_2^4$ \\ \hline                       
\end{tabular}
\caption{$m= 1, d = 2$ bosonic symmorphic crystallographic phases ($H^2_P(\mathbb{T}^2;\mathbb{Z})$) vs $d=2$ bosonic symmorphic crystallographic SPT phases ($H^4(G;\mathbb{Z})$). Input taken from \cite{Gomi-Twists} and \cite{Thorngren-SET}.}
\label{Table1}
\end{table}

It turns out that $H^2_P(\mathbb{T}^2; \mathbb{Z})$ is isomorphic to the second cohomology group of the \emph{reciprocal group} of the total symmetry group $G$, i.e., the group $G^{*}$ generated by the point group together with translations in the reciprocal lattice (see Appendix \ref{sec:Equiv-computations}) and, because of the difference in degree and the difference between $G$ and $G^*$, in general these two should yield different results. However for $\mathsf{pm}$ and $\mathsf{cm}$ the number of phases is the same. For $\mathsf{p2}$, $\mathsf{p3}$, $\mathsf{p4}$, $\mathsf{p6}$ and $\mathsf{p3m1}$ our $m = 1, d= 2$ has more phases than the SPTs found by Thorngren and Else whereas for $\mathsf{pmm}$ and $\mathsf{cmm}$, $\mathsf{p4m}$, $\mathsf{p6m}$ and $\mathsf{p31m}$ there are more SPT phases. Where there is a difference between our results and the SPT results it is clearly due to us having made fewer assumptions: but whether it is solely due to not having imposed short-range entanglement or stability under stacking with the trivial phase, or whether it is due to a true physical difference between emergent relativistic topological field theories and gapped topological phases, remains an interesting open question.

Notice that neither having more nor fewer phases than in the SPT classification should be surprising. For example, if the difference in results were due to us not assuming stability under the so-called stacking operation with a trivial phase, then we may see more phases consisting of fragile crystalline topological phases \cite{Vishwanath-Fragile}, which are robust under adiabatic evolution to interacting systems \cite{Else-Fragile}. On the other hand, if the difference also lies in the short-range entanglement condition, it may also happen that we see fewer phases: two distinct SPT phases, one of which only arises in interacting systems, which by virtue of being distinct are \emph{not} connected by adiabatic evolution through short-range entangled states, could nevertheless be connected by unconstrained adiabatic evolution ---such SPT phases would merge into a single phase in our classification.

We now compare our result with non-interacting phases. It is well known that there are no non-interacting topological phases for bosonic systems. As we shall show in section \ref{sec:Examples}, every non-degenerate interacting fermionic system can be adiabatically evolved to a free fermion system. However, as we show below, not every free fermion topological phase is stable under adiabatically evolving through interactions. On the other hand non-interacting fermionic phases stabilized under stacking are classified by twisted equivariant $K$-theory \cite{Kitaev}, \cite{Freed-Moore}, \cite{Gomi-Wallpaper}. Therefore, our interacting phases are not necessarily a subset of these since we can also capture fragile phases \cite{Vishwanath-Fragile} which are stable under such adiabatic evolution.  We consider the simplest case where $P_F = P\times \mathbb{Z}_{2}^{F}$, which leaves out some interesting purely fermionic phases, and compare $H^{2}_{P\times \mathbb{Z}_{2}^{F}}(\mathbb{T}^{2};\mathbb{Z})$ with those of Shiozaki et al. \cite{Gomi-Wallpaper}, which use the $K$-group $\bar{K}^{0}_{P}(\mathbb{T}^2)$, those being the symmorphic ones in class $A$ \footnote{By $\bar{K}^{0}_P(\mathbb{T}^{d})$ we mean the kernel of the homomorphism $K^{0}_P(\mathbb{T}^{d})\rightarrow K^{0}(\cdot)$. This is because \cite{Gomi-Wallpaper} includes the virtual dimension of the bundle, which has no physical interpretation. This is tantamount to removing a $\mathbb{Z}$ factor.}. As reflections may interact non-trivially with $\mathbb{Z}^{F}_{2}$, we only consider groups without reflections, i.e. $\mathsf{p2}$, $\mathsf{p3}$, $\mathsf{p4}$ and $\mathsf{p6}$ as a more sensible comparison. We compare them in Table \ref{Table2}:

\begin{table}[htp]
\begin{tabular}{cccc}
\hline
{\centering{$G$}} &  $P$ &       $H^2_P(\mathbb{T}^2;\mathbb{Z})$        & $\bar{K}^{0}_{P}(\mathbb{T}^2)$    \\
\hline
        $\mathsf{p2}$   & $\mathbb{Z}_2$            & $\mathbb{Z}\oplus\mathbb{Z}_2^{3}$                     & $\mathbb{Z}^{5}$                   \\
 $\mathsf{p3}$   & $\mathbb{Z}_3$             & $\mathbb{Z}\oplus\mathbb{Z}_3^{2}$                     & $\mathbb{Z}^{7}$                   \\
 $\mathsf{p4}$  & $\mathbb{Z}_4$             & $\mathbb{Z}\oplus\mathbb{Z}_2\oplus\mathbb{Z}_{4}$ & $\mathbb{Z}^8$ \\
 $\mathsf{p6}$   & $\mathbb{Z}_6$             & $\mathbb{Z}\oplus \mathbb{Z}_{6}$                     & $\mathbb{Z}^{9}$                   \\ \hline                       
\end{tabular}
\caption{$m= 1, d = 2$ fermionic symmorphic crystallographic phases ($H^2_P(\mathbb{T}^2;\mathbb{Z})$) without the $\mathbb{Z}_2^{F}$ parity index  vs $d=2$  symmorphic crystallographic non-interacting fermionic phases ($\bar{K}^{0}_{P}(\mathbb{T}^2)$). Symmetry groups are those without reflections i.e. $\mathsf{p2}$, $\mathsf{p3}$, $\mathsf{p4}$ and $\mathsf{p6}$. We have removed the non-physical virtual dimension $\mathbb{Z}$-factor included in \cite{Gomi-Wallpaper}. For these groups there are many more non-interacting phases than interacting ones. Input taken from \cite{Gomi-Twists} and \cite{Gomi-Wallpaper}.}
\label{Table2}
\end{table}
We note that the case $P_F = P\times \mathbb{Z}_{2}^{F}$ for $m = 1$ yields
\begin{equation}
    H^{2}_{P_F}(\mathbb{T}^2;\mathbb{Z}) \approx H^2_P(\mathbb{T}^2;\mathbb{Z}) \oplus \mathbb{Z}_2^{F}
\end{equation}
which essentially gives the same classes as bosons except for the parity factor, which we do not include in Table \ref{Table2}. We find that there are far more non-interacting phases than interacting ones for these groups, indicating that many non-interacting phases can be connected by adiabatic evolution through interacting states.

\section{Explicit models}\label{sec:Examples}
The purpose of the present section is to provide physical examples of systems which have non-zero invariants over the generalized Brillouin zone i.e. which correspond to a non-trivial topological phase in $H^2_{P}(\mathbb{T}^d;\mathbb{Z})$.

\subsection{Warm up, only translations, $d = 2$}
Consider a system of $n$ non-interacting particles (let us ignore fermionic indistinguishability for simplicity) in $d =2$ i.e. the Hamiltonian is of the form:
\begin{equation}\label{eq:Product-Hamiltonian}
    \mathcal{H} = \sum_{i = 1}^{n} \mathcal{L}_{i}\otimes \bigotimes_{j\neq i} I_{j},
\end{equation}
with $\mathcal{L}_{i} = \mathcal{L}$ the same single particle Hamiltonian. Furthermore let us consider a single particle Hamiltonian $\mathcal{L}$ which, by itself, has a non-trivial Chern number over the single particle Brillouin zone, which we denote as $\mathbb{T}^{2}_{i}$, where the $i$ indicates the single particle Brillouin zone corresponding to the $ith$-particle. We shall prove that this non-interacting many-body system has a non-trivial Chern number over the generalized Brillouin zone, namely, it will be the same as the Chern number of the single particle wave function. The ground state of $\mathcal{H}$ is a tensor product of states $\Psi = \bigotimes_{i}^{n} \psi_i $. Let us assume for simplicity that the single particle Hamiltonian $\mathcal{L}$ has only one energy band $E_{1}(\vec{k})$ below the Fermi energy $E_{F}$ with corresponding eigenfunction $\psi_{1}(\vec{k})$ and that our system is gapped. Then we know from the single particle picture that each $\psi_i$ will be a distribution over the $\psi_1$: 
\begin{eqnarray}
    \Psi &\simeq& \bigotimes_{i=1}^{n}\left[\int^{\oplus}_{\mathbb{T}^{2}_{i}} \psi_{1,i}(\vec{k}_{i})d\vec{k}_{i}\right] \nonumber \\
     &\simeq& \int^{\oplus}_{\mathbb{T}^{2n}} \left(\bigotimes_{i=1}^{n} \psi_{1,i}(\vec{k}_{i})\right)\sqcap_{i} d\vec{k}_{i},
\end{eqnarray}
where we have used the distributive property to rewrite the tensor product as a direct integral over the product of the single-particle Brillouin zones.

We now wish to rewrite $\Psi$ in terms of the generalized Bloch decomposition, where each $\vec{k}$ in the generalized Brillouin zone is, in this case, a sum over the $\vec{k}_{i}$'s i.e. $\vec{k} = \sum_{i =1}^{n} \vec{k}_{i}$. So, by rewriting, say,  $\vec{k}_n$ in terms of $\vec{k}$ we can perform the following decomposition:
\begin{eqnarray}
    \Psi &\simeq& \int^{\oplus}_{\mathbb{T}^2}\Psi(\vec{k})d\vec{k}\,,\\
    \Psi(\vec{k}) &=& \int^{\oplus}_{\mathbb{T}^{2(n-1)} }\bigotimes_{i=1}^{n-1}\psi_{1,i}(\vec{k}_i)\otimes \psi_{1,n}\Bigl(\vec{k} - \sum_{i=1}^{n-1}\vec{k}_{i}\Bigr)\sqcap_{i} d \vec{k}_{i}.
\end{eqnarray}
   
We can therefore compute the many-body Chern number of $\Psi(\vec{k})$ using the Berry connection:
\begin{equation}
\begin{split}
       A_{\rho}(\vec{k}) &= -i\int_{\mathbb{B}^{n}}\Psi^{*}(\vec{k})\frac{\partial}{\partial k_{\rho}}\Psi(\vec{k})\sqcap_{j}d\vec{x}_{j}\\
       &= -i\int_{\mathbb{B}^{n}}\int_{\mathbb{T}^{2(n-1)}}\left(\prod_{i=1}^{n-1}|\psi_{1,i}(\vec{k}_i)|^2\right) \psi^{*}_{1,n}\Bigl(\vec{k} - \sum_{i=1}^{n-1}\vec{k}_{i}\Bigr)\frac{\partial}{\partial k_{\rho}}\psi_{1,n}\Bigl(\vec{k} -\sum_{i=1}^{n-1}\vec{k}_{i}\Bigr)\sqcap_{i} d \vec{k}_{i}\sqcap_{j}d\vec{x}_{j}\nonumber
\end{split}
\end{equation}
and curvature
\begin{equation}
\begin{aligned}
        F_{\sigma\rho}(\vec{k}) & = \frac{ \partial A_{\sigma}(\vec{k})}{\partial k_{\rho}} - \frac{ \partial A_{\rho}(\vec{k})}{\partial k_{\sigma}}\\
       & = i\int_{\mathbb{B}^{n}} \frac{\partial \Psi^{*}(\vec{k})}{\partial k_{\sigma}}\frac{ \partial \Psi(\vec{k})}{\partial k_{\rho}} -\frac{\partial \Psi^{*}(\vec{k})}{\partial k_{\rho}}\frac{ \partial \Psi(\vec{k})}{\partial k_{\sigma}} \sqcap_{j}d\vec{x}_{j}\\
       & \begin{split}= i\int_{\mathbb{B}^{n}} \int_{\mathbb{T}^{2(n-1)}}
       &\left(\prod_{i=1}^{n-1}|\psi_{1,i}(\vec{k}_i)|^2\right)\\
       &\left(\frac{\partial}{\partial k_{\sigma}}\psi^{*}_{1,n}\Bigl(\vec{k} - \sum_{i=1}^{n-1}\vec{k}_{i}\Bigr)\frac{\partial}{\partial k_{\rho}}\psi_{1,n}\Bigl(\vec{k} -\sum_{i=1}^{n-1}\vec{k}_{i}\Bigr)\right.\\
       &\quad-\left.\frac{\partial}{\partial k_{\rho}}\psi^{*}_{1,n}\Bigl(\vec{k}- \sum_{i=1}^{n-1}\vec{k}_{i} \Bigr) \frac{\partial}{\partial k_{\sigma}}\psi_{1,n}\Bigl(\vec{k} -\sum_{i=1}^{n-1}\vec{k}_{i}\Bigr)\right)\sqcap_{i} d \vec{k}_{i}\sqcap_{j}d\vec{x}_{j},
       \end{split}\nonumber
\end{aligned}
\end{equation}
with $\mathbb{B}^{n}$ the product of $n$ fundamental regions $\mathbb{B}$. Thus the many-body Chern number is simply:

\begin{equation}
    \sigma_{xy} = \frac{-i}{2\pi}\int_{\mathbb{T}^2}F_{xy}d\vec{k}
\end{equation}

Putting all the integrals in the $\vec{k}$'s together, we have an integral over $\mathbb{T}^{2(n-1)}\times \mathbb{T}^{2}$. The main difficulty in performing the integral are the arguments of the form $\vec{k}-\sum_{i=1}^{n-1}\vec{k}_{i}$. We can perform a trick, using the automorphism $g$ of $\mathbb{T}^{2(n-1)}\times \mathbb{T}^{2}$ which maps
\begin{equation}
    \bigl(\vec{k}_{1},...,\vec{k}_{n-1},\vec{k}\bigr)
    \mapsto \Bigl(\vec{k}_{1},...,\vec{k}_{n-1},\vec{k} - \sum_{i=1}^{n-1}\vec{k}_{i}\Bigr),
\end{equation}
whose Jacobian $Dg$ has $\det Dg = 1$. Thus, our integral is equivalent to one under the image of $g^{-1}$, which transforms the arguments $\vec{k}-\sum_{i=1}^{n-1}\vec{k}_{i}$ into $\vec{k}$. Now we can perform the integral first with respect to the $\vec{k}_{i}$'s and then with respect to $\vec{k}$. The first integral is equal to the product of the norms of the wave functions $\psi_{1,i}$, which are normalized to $1$. The second one is simply equal to the single-particle Chern number $\sigma^{s}_{xy}$, hence 
\begin{equation}
    \sigma_{xy} = \sigma^{s}_{xy}.
\end{equation}

We have thus constructed a model (a non-interacting one) where the many-body Chern number over the generalized Brillouin zone has a physical interpretation and can be non-zero in $H^2(\mathbb{T}^2;\mathbb{Z})$, namely the same one as the Chern number constructed from the single particle Brillouin zone, i.e. the conductance $\sigma^s_{xy}$ of the system and its class in $H^2(\mathbb{T}_i^{2};\mathbb{Z})$. However, we can conclude something stronger. The derivation in the previous sections implies that the many-body Chern number will be the same even if we adiabatically evolve the system via including interaction terms, so long as these do not break the gap nor ground state degeneracy. We conclude something which is already well known: the quantization of the conductance of non-interacting gapped systems is stable under the inclusion of certain types of particle-particle interactions. Which particle-particle interactions do not break the gap nor degeneracy? This has been already addressed in quantitative terms in \cite{Hastings-Interacting}, \cite{Sven-Interacting}.
\subsection{$P = \mathbb{Z}_2\times \mathbb{Z}_2,\, d = 3$}\label{subsec:F222}
Let us consider now a system with a Hamiltonian of the form (\ref{eq:Product-Hamiltonian}) in $d =3$, where each single particle Hamiltonian $\mathcal{L}_i$ has crystalline symmetry group $F222$. The group $F222$ has as point group $P =\mathbb{Z}_2\times \mathbb{Z}_2$, where each factor is generated by rotations of order 2 around the $x$ and $y$ axes. The corresponding lattice is \textit{face-centered}, and has as reciprocal lattice vectors (in the appropriate units) $\vec{b}_1 = (-1,1,1),\, \vec{b}_2 = (1,-1,1),\,\vec{b}_3 = (1,1,-1)$. The model, first presented (to our knowledge) in \cite{Shiozaki-AHSS} and \cite{Shiozaki-SPT}.  The wave-function associated to $\mathcal{L}$ produces a Berry connection $\vec{A}_{s}(\vec{k}_i)$ and curvature $F_s(\vec{k}_i)$ respectively. The fundamental region for the entire crystallographic group is a quarter cube of the single particle Brillouin zone (again see \cite{Shiozaki-AHSS} for a complete sketch). The quotient of this boundary by the $P$-action yields a real projective plane $\mathbb{R}P^2$, and we can see that $\mathcal{L}$ represents a nontrivial phase because we generate the non-zero cohomology class in $H^2(\mathbb{R}P^2; \mathbb{Z}) = \mathbb{Z}_2$ using $\vec{A}_{s}(\vec{k}_i)$ and $F_s(\vec{k}_i)$ in the formula \cite{Freed-Torsion}
\begin{equation}\label{eq:Torsion-RP2}
    c_1 = \frac{i}{2\pi} \ln \Hol_{l}(\vec{A}_s) + \frac{1}{2}\frac{i}{2\pi}\int_{X} F_{s}
\end{equation}
where $\Hol_{l}(\vec{A})$ denotes the holonomy of $\vec{A}$ around the loop $l$, which is the generator of the first homology group $H_1(\mathbb{R}P^2;\mathbb{Z})$, and $X$ is the 2-dimensional cycle attached to it. In terms of the quarter Brillouin zone, $l$ is the cycle made out of 3 of edges after taking the quotient, and similarly, $X$ is made out of the corresponding 3 faces and taking the quotient.
We now come back to the many-body system of $n$ non-interacting particles \ref{eq:Product-Hamiltonian} with a single-particle Hamiltonian as above. Once more, the ground-state will be made of a tensor product of single-particle wave-functions $\Psi = \bigotimes_{i=1}^{n}\psi_i$. We shall repeat a similar procedure to that of our previous example, but now at the level of direct integrals. We can first decompose $\Psi$ into a direct integral over the product of the single-particle Brillouin zones $\mathbb{T}^{3n}$ rewriting the tensor product of direct integral decompositions as a direct integral over $\mathbb{T}^{3n}$. We can then pick the $n$-th wave function and its crystal momentum $\vec{k}_{n}$ and rewrite it as $\vec{k} - \sum_{i=1}^{n-1}\vec{k}_i$, where $\vec{k} = \sum_{i=1}^{n}\vec{k}_i$ is, once more, a point on the generalized Brillouin zone $\mathbb{T}^3$. Now, we rewrite the direct integral over $\mathbb{T}^{3(n-1)}\times \mathbb{T}^3$ as a direct integral over $\mathbb{T}^3$ only, where the integrand itself is the direct integral
\begin{equation}
    \Psi(\vec{k}) = \int_{\mathbb{T}^{3(n-1)}}^{\oplus} \bigotimes_{i}^{n-1}\psi_i(\vec{k}_{i}) \otimes \psi_{n}\Bigl(\vec{k}-\sum_{i=0}^{n-1}\vec{k}_i\Bigr)\sqcap_{i}^{n-1} d \vec{k}_{i}.
\end{equation}
There is a line bundle over $\mathbb{T}^3$ associated to this direct integral which we also denote as $\Psi$ and whose fiber at each $\vec{k}$ is essentially the span of $\Psi(\vec{k})$. We will now show there is a $P$-equivariant vector bundle isomorphism between $\Psi$ and the line bundle we shall call $\Psi^g$, defined to have fibers given by the the span of:
\begin{equation}
    \Psi^{g}(\vec{k}) = \int_{\mathbb{T}^{3(n-1)}}^{\oplus} \bigotimes_{i=1}^{n-1}\psi_i(\vec{k}_{i}) \otimes \psi_{n}(\vec{k})\sqcap_{i}^{n-1} d \vec{k}_{i}
\end{equation}
Indeed, such an isomorphism $\Psi^g \cong \Psi$ is provided by $\int_{\mathbb{T}^3}^\oplus \bigotimes_{i=1}^n \mathrm{Id} \otimes M d\vec{k}$, where $M$ is the operator of multiplication by the function $e^{-i\sum_{i=1}\vec{k}_{i}\cdot \vec{x}_{n}}$, because by Bloch's theorem, $\psi_n$ satisfies:
\begin{equation}
\psi_n\Bigl(\vec{k}-\sum_{i=1}^{n-1}\vec{k}_i\Bigr)(\vec{x}_n) = e^{-i\sum_{i=1}\vec{k}_{i}\cdot \vec{x}_{n}}\psi_{n}(\vec{k})(\vec{x}_{n}).
\end{equation}

The difference between $\Psi^{g}$ and $\Psi$ is simply that the last factor in the tensor product is now independent of the $\vec{k}_{i}$'s and thus the Berry connection and curvature $\vec{A}_g$ and $F_{g}$ of $\Psi^{g}$ over the generalized Brillouin zone $\mathbb{T}^3$ are equivalent to the single-particle ones $\vec{A}_s$ and $F_s$ associated to our single particle Hamiltonian $\mathcal{L}$ over the single particle Brillouin zone $\mathbb{T}^3_i$ since the integrals over the $\vec{k}_{i}$'s equal 1 in the definitions of $\vec{A}_g$ and $F_{g}$, as in the previous subsection. Using $\vec{A}_g$ and $F_{g}$ in the formula (\ref{eq:Torsion-RP2}) and looking at the same fundamental region but for the generalized Brillouin zone, will yield a non-trivial class and since $\Psi^{g}$ and $\Psi$ are isomorphic, we conclude that $\Psi$ represents a non-trivial phase in $H^2_{\mathbb{Z}_2\times \mathbb{Z}_2}(\mathbb{T}^3;\mathbb{Z})$. Which class in $H^2_{\mathbb{Z}_2\times \mathbb{Z}_2}(\mathbb{T}^3;\mathbb{Z})$ is it? The calculation for the whole group yields:
\begin{equation}
    H^2_{\mathbb{Z}_2\times \mathbb{Z}_2}(\mathbb{T}^3;\mathbb{Z}) = \mathbb{Z}_{4}\oplus \mathbb{Z}_2^2
\end{equation}

Indeed, a rational cohomology calculation shows that the group is torsion, so the universal coefficients theorem says it is $H_1((\mathbb{T}^3 \times EP)/P; \mathbb{Z})$, which is the abelianization of $\pi_1((\mathbb{T}^3 \times EP)/P) \cong \mathbb{Z}^3 \rtimes (\mathbb{Z}_2 \times \mathbb{Z}_2)$.
Thus the $\mathbb{Z}_4$ summand actually comes from the action of $\mathbb{Z}_2\times \mathbb{Z}_2$ on the fundamental group of $\mathbb{T}^3$, as did our invariant. To see what this $\mathbb{Z}_2$ invariant corresponds to in $H^2_{\mathbb{Z}_2\times \mathbb{Z}_2}(\mathbb{T}^3;\mathbb{Z})$ consider that the copy of $\mathbb{RP}^2$ used to define the invariant sits inside of $\mathbb{T}^3/P \cong \mathbb{RP}^3$. The inclusion of $\mathbb{RP}^2$ in $\mathbb{RP}^3$ induces an isomorphism in $H^2$, and the canonical map $(\mathbb{T}^3 \times EP)/P \to \mathbb{T}^3/P$ can be shown to induce on $H^2$ the inclusion of $\mathbb{Z}_2$ as the order 2 subgroup of the $\mathbb{Z}_4$ summand.

It was argued in \cite{Shiozaki-SPT} that this non-interacting phase extended to a $d=3$ fermionic SPT phase. Hence, this example shows that systems representing a non-trivial phase in our derivation can also represent a non-trivial SPT phase.

\subsection{Stability of free fermion phases with a single valence band under interacting adiabatic evolution}

The above examples are evidence that as long as one has a fermionic non-interacting system with a single band below the Fermi energy, one can use our direct integral decomposition over the $dn$-dimensional torus and our automorphism $g$ defined above to construct an invariant which is stable under adiabatic evolution to systems with interaction terms, so long as the evolution does not break the symmetry, gap nor non-degeneracy. Moreover, there is a one-to-one correspondence between phases of single-particle fermionic systems with a unique valence band and phases of many-body non-degenerate fermionic systems, simply exchanging the single-particle Brillouin zone for its many-body analogue: the generalized Brillouin zone. Thus, if one agrees with the mathematical interpretations of our physical assumptions (and as we have explained we are indeed using the same interpretations underlying all available notions of phases in the physics literature) it implies that the topological invariants of these non-interacting fermionic systems and their corresponding physical interpretations are robust to adiabatic evolution through any types of interactions that respect the above mentioned constraints. Which interactions do not break the symmetry, gap nor non-degeneracy? We hope to have a quantitative answer in the spirit of \cite{Hastings-Interacting}, \cite{Sven-Interacting} in future work. We further emphasize that, as mentioned above, these are not all of the non-interacting phases, but only those which arise when the associated single particle vector bundle is a line bundle. This means that all other non-interacting fermionic phases can be connected by an interacting adiabatic path to one of these non-interacting phases that arise from single valence bands. Note that we did not assume stability of these phases under stacking with the trivial phase, hence we are also including the fragile ones \cite{Vishwanath-Fragile}, some of which have been shown to be stable under interactions through field theory methods \cite{Else-Fragile} and here we are deducing that only those arising from $P$-equivariant single particle line bundles are stable under adiabatic evolution through all interactions that preserve symmetry, gap and non-degeneracy. Finally, let us remark that nothing analogous can happen for degenerate systems nor for bosons as all of these must be, strictly speaking, made from interacting systems.

\section{$m >1$, the degenerate case}
We now have $\mathsf{Vect}^{m}_{P}(\mathbb{T}^{d})$ and $\mathsf{Vect}^{m}_{P_{F}}(\mathbb{T}^{d})$ as the full set of $m$-degenerate symmorphic topological phases. Note that this case can only arise in systems with interactions. Mathematically these objects are not as well understood as the $m =1$ case since maps to $BGL(m)$ are no longer given by some generalized cohomology group. One could, analogously to the non-interacting case, consider stabilizing the dimension of the bundles and obtaining $P$-equivariant $K$-theory, however it comes at a non-trivial price for we would be stabilizing the ground state degeneracy and further analysis is required to see how much physics remains after such an operation. But we could still obtain a partial classification using equivariant Chern characteristic classes \cite{Bott-Tu}
$c_i (\mathcal{H})\in H^{2i}_{P}(\mathbb{T}^d; \mathbb{Z})$. Phases with different equivariant Chern classes are different, whereas phases with the same Chern classes may still be different, but perhaps can only be distinguished through a finer invariant. Similarly to the non-degenerate case, we do not know what stacking corresponds to in terms of operations with these cohomology classes. One might be tempted to think that stacking is a tensor product of the associated vector bundles, however the direct integral decomposition of a tensor product is not the fiberwise tensor product but rather what is known as an \textit{external} tensor product \cite{Hatcher-K-theory}, and this is without taking into account the bosonic and fermionic nature of the wave-functions, which most likely modifies the external tensor product operation in a non-trivial manner.

\subsection{$C_2$ Bosonic Examples}
The simplest examples are those for bosons where $d =2$ and $P = C_2$ , the group with 2 elements. There are 3 symmorphic crystallographic groups, of which for brevity we will only discuss two, $\mathsf{pm}$ (reflections over a fixed parallel axis) and $\mathsf{cm}$ (glide reflections) \cite{Gomi-Twists}.
A way to synthesize all cohomology groups $H^{i}_{P}(\mathbb{T}^d;\mathbb{Z})$ is by giving the full cohomology ring $H^{*}_{P}(\mathbb{T}^d;\mathbb{Z})$. For $\mathsf{pm}$, the action on the generalized Brillouin zone is $(k_1,k_2)\mapsto (k_1,-k_2)$ and we can compute (see Appendix \ref{sec:Equiv-computations}):
\begin{equation}\label{eq:pm}
    H^{*}_{C_2}(\mathbb{T}^{2};\mathbb{Z}) \cong \mathbb{Z}[x,y,t]/(t^2,2x,2y,xy),
\end{equation}
where $t$ has degree 1 and $x$ and $y$ have degree 2.
On the other hand for $\mathsf{cm}$, the action is $(k_1,k_2)\mapsto (k_2,k_1)$ and yields
\begin{equation}\label{eq:cm}
    H^{*}_{C_2}(\mathbb{T}^{2};\mathbb{Z}) \cong \mathbb{Z}[t,u,v]/(2t,2v,u^2,ut,uv,v^2),
\end{equation}
where $u$, $t$ and $v$, have degrees $1$, $2$ and $3$, respectively. We note that equivariant characteristic classes live in these cohomology rings, however, which cohomology classes are characteristic classes remains an open problem.

\section{Differences with SPT and SET}\label{sec:Difference}
Besides our lack of relativistic fields and spectra, let us discuss some of the differences between the assumptions made in the constructive approaches and ours. SPT phases are generally taken to have a unique ground state, just as our non-degenerate phases, but said ground state should be short-range entangled \cite{Wen-SPT}. We found that all systems representing our non-degenerate interacting fermionic phases can be connected adiabatically to a non-interacting free fermion system. Does it not contradict our finding that we can have more of these phases than SPTs according to the results of Thorngren-Else \cite{Thorngren-SET}? At first glance it could seem contradictory since our derivation should include all SPT paths and free fermions are systems in a given SPT phase, so should there not be a path connecting two apparently distinct interacting phases adiabatically connected to non-interacting ones but which seem to be the in the same SPT? SPT classifications also assume stability of phases under stacking with the trivial phase. We have not put in any such criteria and, by the results of section \ref{sec:Examples}, we know that for fermions any non-trivial fragile phase arising from single band (line bundle) in non-interacting systems will be stable under adiabatic evolution through interactions which respect the non-degeneracy, symmetry and gap. Thus our phases also include ones which are not stable under the stacking operation, implicitly unaccounted in all SPT constructions and hence we are seeing all non-interacting fragile phases which are stable to interactions \cite{Else-Fragile}. In our approach systems can transition adiabatically from short-range entangled states (in fact, for fermions, from product states) to long-range entangled states and back, so long as the ground state remains unique. Thus, our non-degenerate phases are in principle more robust than crystallographic SPTs. What happens in cases where there are fewer crystalline phases than SPTs? In the case where the difference is due to short-range entanglement it means that some of these SPT are not stable under the inclusion of long-range interactions. However, the $F222$ example in subsection \ref{subsec:F222} constitutes a fermionic crystalline SPT phase which is stable under including long-range interactions adiabatically.

There are often further restrictions on the allowed actions of $P$ as it is assumed that certain subspaces should be left invariant \cite{Else-Nayak-SPT}. Here we did not put any restriction on the type of $P$-actions, except that they arise from crystallographic groups. Another difference is that some of the constructive approaches build their invariants through boundary modes, whereas we assumed our systems had no boundary.  As we mentioned in the introduction these results do not imply that published SPT classifications are incorrect since we start with a subset of their assumptions. SET phases on the other hand construct their classification based on the fractionalization of quasi-particle excitations \cite{Cheng-SET} such as anyons. We only employed the ground state bundle and never addressed excitations. In fact we did not even assume the existence of quasi-particles, hence our $m$-degenerate topological phases could in principle occur in strongly interacting systems that have no quasi-particles. However, in spin systems, Lieb-Schultz-Mattis type theorems \cite{Hastings-lieb} for gapped systems satisfying a Lieb-Robinson bound forces many of them to have quasi-particle excitations and thus, they may be generally forced upon us.

\section{Conclusions}\label{sec:Conclusions}
We have derived from first principles and a generalized Bloch transform a classification for crystalline symmorphic interacting gapped phases without assuming stability under stacking nor short-range entanglement. We have seen that for non-degenerate fermionic ground states the classification is complete, is given by $H^2_{P}(\mathbb{T}^d;\mathbb{Z})$ and each phase has a non-interacting representative, i.e. systems in each non-degenerate phase can be connected adiabatically to a non-interacting system. Not all non-trivial non-interacting fermionic phases represent non-trivial interacting ones. We also showed an example of such a phase in $d=3$ with symmetry group $F222$ which is both one of our non-degenerate interacting phases and an SPT phase. We further show how, for the degenerate case, one can employ characteristic classes and cohomology to achieve a partial classification and computed some examples. In future work we hope to be able to model the short-range entanglement condition mathematically (but not the emergent conditions), so that we can obtain a classification of phases for short-range entangled states and adiabatic evolution through them. This would help to see if the discrepancies observed between our classification and existing ones are solely due to our inclusion of crystalline phases which either are not stable under stacking and are destroyed when allowing long-range entangled states or whether the other assumptions (emergent relativistic field theory/spectrum) are responsible. Independently of either scenario, our results stand on their own as these new phases can be viewed as crystalline topological phases that do not require these extra conditions to be satisfied, and are hence more robust in that sense.

Finally we remark that lattice systems with on-site symmetry given by a compact Lie group $G$, such as an $SU(2)$ spin chain, the Toric code or Cubic Haah's code, are also translation invariant and we expect to have something similar to $H^{2}_{G}(\mathbb{T}^d;\mathbb{Z})$. However they are examples that have projective representations and time-reversal symmetry \cite{Cheng-SET}, which adds a real structure \cite{D-G-AI} to the ground state bundle and are outside our current scope. We will explore this in future work.

\acknowledgments{We would like to thank the referee for providing critiques which greatly improved the overall quality of our manuscript. We also thank Ken Shiozaki for providing us with reference \cite{Shiozaki-SPT} and Arun Debray for useful comments. D. Sheinbaum acknowledges funding from CONACYT Frontera 42821.}

\appendix
\section{\label{sec:TranslationA}Translation and Fock space decomposition}
In this section we follow very closely the work of Gomi \cite{Gomi-Twists}, with slight modifications. Let $L^{2}(\mathbb{R}^{d};W)$ be the complex separable Hilbert space of square integrable functions on $\mathbb{R}^d$ with values in an $SU(2)$-representation $W$. $L^{2}(\mathbb{R}^{d};W)$ is the Hilbert space of a single particle in $\mathbb{R}^{d}$ and the dimension of $W$ determines whether the particle is a boson or a fermion. The $n$-th tensor product of our Hilbert space $L^2(\mathbb{R}^{d};W)^{\otimes n}$ is isomorphic to $L^2(\mathbb{R}^{nd};W^{\otimes n})$. There is a canonical action of the group $\mathbb{Z}^d$ on $\mathbb{R}^d$ by discrete translations and hence there is an induced action on the $n$ copies of $\mathbb{R}^{d}$ in $\mathbb{R}^{nd}$. This is simply translating each element of $\mathbb{R}^{d}$ by a fixed vector in $\mathbb{Z}^d$. We denote this action on $\mathbb{R}^{nd}$ by $\vec{a}\cdot x,\,\vec{a} \in \mathbb{Z}^{d},\, x\in \mathbb{R}^{nd}$. Hence there is a canonical action on $L^2(\mathbb{R}^{nd};W^{\otimes n})$, where the new function is obtained by evaluating the old function at the translated point. Here we shall slightly generalize the presentation in \cite{Gomi-Twists} of the Bloch transform using this action on $\mathbb{R}^{nd}$. Let $\widehat{\mathbb{Z}}^{d} = Hom(\mathbb{Z}^{d}, U(1))$ be the Pontryagin dual of $\mathbb{Z}^{d}$. Let us define an intermediate Hilbert space just as in \cite{Gomi-Twists}
\begin{equation}
\begin{split}
    L^{2}_{\mathbb{Z}^{d}}(\widehat{\mathbb{Z}}^{d}\times \mathbb{R}^{nd}; W^{\otimes n})
    = \{ \psi \in  L^2(\widehat{\mathbb{Z}}^{d}\times \mathbb{R}^{nd}; W^{\otimes n})|\\
     \psi(\vec{k}, \vec{a}\cdot x) = e^{i\vec{k}\cdot
    \vec{a}}\psi(\vec{k},x) \}
    \end{split}
\end{equation}

We now define $\mathfrak{B}$ and its inverse $\mathfrak{B}^{*}$ as
\begin{equation}
\begin{split}
    \mathfrak{B}: L^{2}_{\mathbb{Z}^{d}}(\widehat{\mathbb{Z}}^{d}\times \mathbb{R}^{nd}; W^{\otimes n}) \rightarrow L^{2}(\mathbb{R}^{nd};W^{\otimes n})\\
    (\mathfrak{B}\psi^{*})(x) = \int_{\widehat{\mathbb{Z}^{d}}} \psi^{*}(\vec{k},x)d\vec{k}
    \end{split}
\end{equation}

\begin{equation}
\begin{split}
    \mathfrak{B}^{*}: L^{2}(\mathbb{R}^{nd};W^{\otimes n}) \rightarrow L^{2}_{\mathbb{Z}^{d}}(\widehat{\mathbb{Z}}^{d}\times \mathbb{R}^{nd}; W^{\otimes n})\\
    (\mathfrak{B}^{*}\psi)(\vec{k},x) = \sum_{\vec{a}\in\mathbb{Z}^{d}} e^{-i\vec{k}\cdot \vec{a}} \psi(\vec{a}\cdot x)
    \end{split}
\end{equation}
$L^{2}_{\mathbb{Z}^{d}}(\widehat{\mathbb{Z}}^{d}\times \mathbb{R}^{nd}; W^{\otimes n})$ is isomorphic to $L^2$-sections of the bundle $E$ over $\widehat{\mathbb{Z}}^{d}$ defined by
\begin{equation}
   E =  \bigcup_{\vec{k}\in \widehat{\mathbb{Z}}^{d}} L^{2}(\mathbb{R}^{nd}/\mathbb{Z}^{d}; \mathcal{L}|_{\vec{k}\times\mathbb{R}^{nd}/\mathbb{Z}^{d}})\otimes W^{\otimes n}
\end{equation}
where $\mathcal{L}$ is the Poincar\'e line bundle $\mathcal{L} \rightarrow \widehat{\mathbb{Z}}^{d}\times \mathbb{R}^{nd}/\mathbb{Z}^{d} $ built from the quotient identification
\begin{equation}
\begin{split}
     \mathbb{Z}^{d}\times(\widehat{\mathbb{Z}}^{d}\times \mathbb{R}^{nd}\times \mathbb{C}) \rightarrow \widehat{\mathbb{Z}}^{d}\times \mathbb{R}^{nd}\times \mathbb{C} \\
     (\vec{a},\vec{k},x,z) \mapsto (\vec{k},\vec{a}\cdot x, e^{i\vec{k}\cdot \vec{a}}z)
\end{split}
\end{equation}
If we rename $L^{2}(\mathbb{R}^{nd}/\mathbb{Z}^{d}; \mathcal{L}|_{\vec{k}\times\mathbb{R}^{nd}/\mathbb{Z}^{d}})\otimes W^{\otimes n}$ as $\mathfrak{H}^{n}(\vec{k})$ we have the equivalence to the direct integral decomposition
 \begin{equation}
 L^2(\mathbb{R}^{dn};W^{\otimes n}) \cong \int^{\oplus}_{\widehat{\mathbb{Z}}^d} \mathfrak{H}^{n}(\vec{k})d\vec{k}\,,
\end{equation}

Fock space is defined as the subspace of square integrable functions of $\bigoplus_{n\geq 0}L^2(\mathbb{R}^{dn};W^{\otimes n})$. So rewriting each $L^2(\mathbb{R}^{dn};W^{\otimes n})$ we have 

\begin{equation}\label{eq:Fock-direct2}
    \mathcal{F}(L^2(\mathbb{R}^{d};W))\cong \int^{\oplus}_{\mathbb{T}^{d}} \bigoplus_{n\geq 0}\mathfrak{H}^{n}(\vec{k})d\vec{k}\,.
\end{equation}

This is simply a rewriting of Fock space. Note that we can restrict to symmetric and antisymmetric Fock spaces and repeat the split, so it works the same for bosons and fermions.

\section{The non-degenerate case leads to Borel cohomology}\label{sec:Borel}

We explained in the main text that phases of interacting bosonic systems with a symmorphic crystallographic symmetry group with point group $P$ are classified by $P$-equivariant rank $m$ vector bundles on $\mathbb{T}^d$ or equivalently by $P$-equivariant homotopy classes of maps $\mathbb{T}^d \to \mathcal{G}_m(\mathfrak{H}')$ from the torus to the Grasmmanian of $m$-dimensional subspaces of the integrand $\mathfrak{H}'$ in the Bloch decomposition of the appropriate Fock space.

In the non-degenerate case, $m=1$, this set of equivariant homotopy classes simplifies to the Borel cohomology group $H_P^2(\mathbb{T}^d; \mathbb{Z})$. We mentioned the intuition for this in the main text: that non-equivariantly the Grassmanian is a $K(\mathbb{Z},2)$ space which classifies $H^2$. But there is a subtlety in relation to the $P$ action as we now explain. The Grassmanian $\mathcal{G}_1(\mathfrak{H}')$ with its $P$-action is a classifying space for $P$-equivariant rank 1 vector bundles. There is well-known equivalence between rank 1 vector bundles, i.e., line bundles, and principal $U(1)$-bundles. Because of this equivalence, $\mathcal{G}_1(\mathfrak{H}')$ can also be thought of as a classifying space for $P$-equivariant principle $U(1)$-bundles. Because $U(1)$ is an Abelian compact Lie group, this means that $\mathcal{G}_1(\mathfrak{H}')$  is $P$-equivariantly homotopy equivalent to the mapping space $\mathrm{Map}(EP, BU(1))$ \cite{LashofMaySegal} and therefore,
\begin{align*}
   [\mathbb{T}^d, \mathcal{G}_1(\mathfrak{H}')]_P & \cong [\mathbb{T}^d, \mathrm{Map}(EP, BU(1))]_P \\ & \cong [\mathbb{T}^d \times EP, BU(1)]_P \\ & \cong[(\mathbb{T}^d \times EP)/P, BU(1)]  \\ & \cong H^2((\mathbb{T}^d \times EP)/P; \mathbb{Z}). 
\end{align*}
The last isomorphism is because $BU(1)$ is a $K(\mathbb{Z},2)$ and that last group is by definition the Borel cohomology group $H_P^2(\mathbb{T}^d; \mathbb{Z})$.

This argument relies heavily on the theorem of Lashof--May--Segal \cite{LashofMaySegal} for compact abelian Lie groups like $U(1)$. In particular, that theorem is not available for $U(m)$ when the degeneracy satisfies $m>1$.

Everything in this section also goes for the fermionic setting (again only in the symmorphic case) simply replacing $P$ with $P_F$ and $G$ with $G_F$.

\section{Equivariant cohomology rings of the 2-torus with $C_2$ point group}\label{sec:Equiv-computations}

To adiabatic evolution classes of non-degenerate bosonic systems with a symmorphic crystallographic symmetry group $G$ with point group $P$ we assigned an element of $H_P^2(\mathbb{T}^d;\mathbb{Z})$, and explained that it can help to know the entire cohomology ring $H_P^*(\mathbb{T}^d; \mathbb{Z})$. Here we include sample calculations for the simplest cases of interest, namely $d=2$ and $P = C_2$. There are three symmorphic plane crystallographic groups with point group of order 2, but we present only two: $\mathsf{pm}$ which contains reflections in lines parallel to one of the translation axes and $\mathsf{cm}$ which contains reflections in line parallel to the bisector of the translation axes.

In both cases the symmetry group is a semi-direct product $(\mathbb{Z} \times \mathbb{Z}) \rtimes C_2$; what differs in each case is the action of $C_2$ on $\mathbb{Z} \times \mathbb{Z}$. The generator $\sigma$ of $C_2$ acts as follows: for $\mathsf{pm}$, $\sigma(x,y) = (x,-y)$ and for $\mathsf{cm}$, $\sigma(x,y) = (y,x)$. From those actions we can compute the action of $P=C_2$ on the Brillouin zone $\mathbb{T}^2 = \mathrm{Hom}(\mathbb{Z}^2, U(1))$. Identifying $\mathrm{Hom}(\mathbb{Z}^2, U(1))$ with $U(1) \times U(1)$ via the map $f \in \mathrm{Hom}(\mathbb{Z}^2, U(1)) \mapsto (f(1,0), f(0,1))$, we can compute the actions on $\mathbb{T}^2$: for $\mathsf{pm}$, $\sigma(w,z) = (w,z^{-1})$ and for $\mathsf{cm}$, $\sigma(w,z) = (z,w)$.

By definition Borel $P$-equivariant cohomology of $\mathbb{T}^2$ is given by ordinary cohomology of the Borel construction, $(\mathbb{T}^2 \times EP)/P$, where $EP$ is the total space of the universal principal $P$-bundle, a contractible space with a free $P$ action. In the case of a torus there is an important simplification: $\mathbb{T}^2$ is the classifying space of the group $\pi_1(\mathbb{T}^2) \cong \mathbb{Z} \times \mathbb{Z}$, and in that case the Borel construction is the classifying space of the group $\pi_1(\mathbb{T}^2) \rtimes P$. With the action of $P=C_2$ on $\mathbb{T}^2$ we can compute the induced action on $\mathbb{Z} \times \mathbb{Z}$ and it turns out to be the same action that defined $G$ as a semi-direct product. Thus $(\mathbb{T}^2 \times EP)/P \simeq BG$.

It is tempting to think that this argument goes through in every symmorphic case, showing that $(\mathbb{T}^d \times EP)/P \simeq BG$, but there is a subtlety in cases when the Bravais lattice is not just $\mathbb{Z}^d$, namely that, while the symmetry group is generated by $P$ and translations by lattice vectors, the Brillouin zone is the quotient of $\mathbb{R}^d$ by the \emph{reciprocal lattice}. The argument does go through as far as showing that the Borel construction is the classifying space of the semi-direct product $\pi_1(\mathbb{T}^d) \rtimes P$, but $\pi_1(\mathbb{T}^d)$ is really the reciprocal lattice, and it is on this lattice that we must consider the action of $P$. If we define the \emph{reciprocal symmetry group} $G^*$ to be the semi-direct product of the reciprocal lattice with $P$, then we obtain that $(\mathbb{T}^d \times EP)/P \simeq BG^*$. In the example above the lattice was $\mathbb{Z}^2$, which is its own reciprocal, so $G \cong G^*$, but in general one must be aware of the distinction. For example, by drawing the Bravais lattice and its reciprocal it is straighforward to see that the symmetry groups $\mathsf{p3m1}$ and $\mathsf{p31m}$ are reciprocal to each other in this sense. But perhaps exactly which group is involved is not as important as the fact that the Borel construction is the classifying of \emph{some} readily computed group, which puts the required equivariant cohomology calculations in the realm of group cohomology for which there is a vast literature.

\emph{Case $G = \mathsf{pm}$}. In this case the group splits as a direct product: $G \cong D_\infty \times \mathbb{Z}$ where the $D_\infty$ factor is generated by the reflection and a translation in a perpendicular direction, and the other factor generated by translation parallel to the axis of reflection. The group $D_\infty$ is isomorphic to the free product $C_2 * C_2$, so we get 
\begin{align*}BG \simeq B((C_2 * C_2) \times \mathbb{Z}) & \simeq (BC_2 \vee BC_2) \times B\mathbb{Z} \\ & \simeq (\mathbb{RP}^\infty \vee \mathbb{RP}^\infty) \times S^1.
\end{align*}
Thus $H^*(BG;\mathbb{Z}) \cong \mathbb{Z}[x,y,t]/(t^2,2x,2y,xy)$ where $t$ has degree 1 and is the generator of $H^*(S^1;\mathbb{Z})$ and $x$ and $y$ have degree 2 and each generate one copy of $H^*(\mathbb{RP}^\infty; \mathbb{Z})$.

\emph{Case $G = \mathsf{cm}$}. For semi-direct products one can always attempt to use of the Lyndon--Hochschild--Serre spectral sequence, which in this case has $E_2^{p,q} = H^p(BC_2; \mathcal{H}^q(\mathbb{T}^2; \mathbb{Z}))$ and converges to $H^{p+q}(BG; \mathbb{Z})$. Here $\mathcal{H}^q(\mathbb{T}^2; \mathbb{Z})$ denotes the $C_2$-module given by $H^q(\mathbb{T}^2; \mathbb{Z})$ with the action of $C_2$ induced by its action on $\mathbb{T}^2$. These are the following $C_2$-modules: for $q=0$, the trivial module $\mathbb{Z}$; for $q=1$, the module $\mathbb{Z} \oplus \mathbb{Z}$ with the $C_2$ action that swaps the summands; and for $q=2$, the module $\mathbb{Z}$ with the sign action. The cohomology of those modules can be readily computed by standard techniques in group cohomology \cite{Adem-Groupcoho}. For $q=0$ it is the integral cohomology ring of $\mathbb{RP}^\infty$, namely, $\mathbb{Z}[t]/(2t)$ where $t$ has bidegree $(2,0)$. For $q=1$, it is concentrated in degree $p=0$ where it is $\mathbb{Z}$ generated by, say, $u$. Finally, for $q=2$, as a module over $H^*(\mathbb{RP}^\infty)$, the cohomology is freely generated by a single element $v$ with bidegree $(1,2)$.

Assembling those results, we see that the $E_2$-page of the spectral sequence is given by $\mathbb{Z}[t,u,v]/(2t,2v,u^2,ut,uv,v^2)$. For a semi-direct product the spectral sequence always collapses at the $E_2$-page, so this is also the $E_\infty$-page. Luckily in each diagonal $p+q=n$ there is exactly one non-zero entry, so there are no additive extension problems, and we have found the cohomology groups of $G$. In fact, the ring structure is also the one given above, but this does not follow solely from the spectral sequence. Simple algebraic considerations on the $E_\infty$-page do show that $ut$, $u^2$, $uv$ are $0$, but as far as the spectral sequence can tell $v^2$ might be either $0$ or $t^3$.

To show that in fact $v^2=0$, we can appeal to a different space where the computation is easier. Consider the map $q : \mathbb{T}^2 \to S^2$ which collapses $\mathbb{T}^2$ to a point. This can be made $C_2$-equivariant by equipping $S^2$ with the $C_2$ action given by reflection in the great circle which is the image under $q$ of the diagonal of $\mathbb{T}^2 = S^1 \times S^1$. Applying the Borel construction to $q$ produces a map $\bar{q} : BG \to (S^2 \times EP)/P$. The $C_2$-action on $S^2$ fixes a great circle pointwise and swaps the two hemispheres it delimits. From this it is straightforward to compute that $(S^2 \times EP)/P \simeq (S^1 \times \mathbb{RP}^\infty)/(S^1 \times \{x_0\}) \cong S^1_+ \wedge \mathbb{RP}^\infty$. The cohomology ring of the latter space is easy to compute with K\"unneth's theorem and in particular there is a class in degree $3$ squaring to $0$ that $\bar{q}^* : H^*(S^1_+ \wedge \mathbb{RP}^\infty; \mathbb{Z}) \to H^*(BG;\mathbb{Z})$ maps to $v$, as required.

\section{Adiabatic evolution vs Chen, Gu and Wen's Local Unitary Evolution and LRE-SRE crossover}\label{sec:LUE}
We have argued that, in principle, non-degenerate systems in the same phase can transition between short-range and long-range entanglement as we never imposed any condition on the type of entanglement systems could have and this could, again in principle, explain the difference between the traditional SPTs and our classification. Another possibility is the relativistic or topological spectrum assumptions. However, this transition between short-range entanglement and long-range entanglement seems forbidden in the notion of phase given by Chen, Gu and Wen \cite{Wen-LUE}. The reason seems to be that the authors substitute the general notion of adiabatic evolution for what they coined local unitary evolution. The main distinction between the the two notions is that in a local unitary evolution a Hamiltonian must be of the form

\begin{equation}\label{eq:Local}
    \mathcal{H} = \sum_{i} \mathcal{O}_{i}
\end{equation}
 with $\mathcal{O}_{i}$ being local. The notion of local here is not entirely precise but intuitively it is that the spatial decay of the norm of these operators will go to zero as the distance is comparable to the entire system's size. With such a definition it indeed seems impossible to go from a short-range entangled system to a long-range one. Chen, Gu and Wen \cite{Wen-LUE} state the reason they prefer this definition is because of practical purposes, as it is much simpler to check whether two systems can be connected by a local unitary evolution than an adiabatic one. We argue there are physically interesting cases which do not fit Chen, Gu and Wen's notion of phase and provide the following as an example of interest. The original explanation of the integer quantum Hall effect is for the non-interacting case (and hence the Hamiltonian is of the form (\ref{eq:Local})). Including Coulomb interactions (non-local terms) opens up new phases (fractional plateaus), nevertheless, for a suitable range of parameters the interactions yield the same value for the Hall conductance as the non-interacting case. This means that we can include non-local interaction terms without breaking the gap, having no critical phenomena and the same value of the Hall conductance. In the local unitary evolution picture these two systems would have to be in a different phase as in one of them the Hamiltonian has these non-local terms. Hence, we choose to include such cases to our notion of phase. This is also an example of a system with short-range entanglement being in the same phase as one with some long-range entanglement. Thus an instantiation of the phenomena which may explain the discrepancy between our results and those of standard SPT constructions.
 
\section{Dimension and Disorder}\label{sec:Disorder}

For non-degenerate ground states we obtained the cohomology group $H^2_{P}(\mathbb{T}^{d};\mathbb{Z})$ as a full classification. The way that the dimension $d$ of a system enters is only through the generalized Brillouin torus $\mathbb{T}^{d}$. Let us remember that the Brillouin zone arises as the Pontryagin dual of the group of translations, meaning that without translation symmetry there is no distinction in dimension. Thus, if we remove translation symmetry completely but keep a symmetry $G'$ which is independent of the dimension (say time-reversal symmetry) we would have something  of the form
\begin{equation}\label{eq:notranslation}
    [*, \mathcal{G}_{m}(\mathfrak{F}(L^2(\mathbb{R};W)))]_{G'}.
\end{equation}
For the non-degenerate case we get $H^2_{G'}(*;\mathbb{Z})$. Equation (\ref{eq:notranslation}) is in stark contrast with results on SPT phases, where the dimensionality of the system is usually reflected in the dimension of the cohomology group $H^{d+1}_{G'}(*;U(1))$ in \cite{Wen-SPT} for example. This discrepancy could be surmounted (perhaps) by the the fact that SPT phases are often further assumed to have a restricted set of actions that leave fixed some chosen subspaces \cite{Else-Nayak-SPT}. To include disorder it would be more reasonable to turn the generalized Brilluoin zone into a noncommutative $C^{*}$-algebra as in the non-interacting case \cite{Prodan-Schulz-Baldes}, \cite{Thiang1} instead of completely removing the Brillouin zone from the picture.

\providecommand{\noopsort}[1]{}\providecommand{\singleletter}[1]{#1}%

\end{document}